\documentstyle[12pt,aps,prc,epsfig,preprint]{revtex}
\tightenlines
\begin{document}

\begin{titlepage}
\title{\vspace*{10mm}\bf Narrow deeply bound $K^-$ atomic states}
\vspace{6pt}

\author{ E.~Friedman and A.~Gal \\
{\it Racah Institute of Physics, The Hebrew University, Jerusalem 91904,
Israel\\}}

\vspace{4pt}
\maketitle

\begin{abstract}
Using  optical potentials fitted to a comprehensive set of strong interaction
level shifts and widths in $K^-$ atoms,
we predict that the $K^-$ atomic levels which are inaccessible in
the atomic cascade process are generally narrow, spanning a range of
widths about 50 - 1500 keV
 over the entire periodic table. The mechanism for this
narrowing is different from the mechanism for narrowing of pionic
atom levels.
Examples of such `deeply bound' $K^-$ atomic states are given,
showing that in many cases these states should be reasonably well resolved.
Several reactions which could be used to form these `deeply bound' states are
mentioned. Narrow deeply bound states are expected also in $\overline{p}$
atoms.
\newline \newline
$PACS$: 21.65.+f; 36.10.Gv
\newline
{\it Keywords}: Kaonic atoms; Deeply bound kaonic states;
$K^-$ nucleus potential \newline\newline
%\vspace{1cm}
Corresponding author: E. Friedman,\newline
Tel: +972 2 658 4667,
Fax: +972 2 658 6347, \newline
E mail: elifried@vms.huji.ac.il
\end{abstract}
\centerline{\today}
\end{titlepage}

The strong interaction level shifts and widths in hadronic atoms
provide valuable information on the hadron-nucleus interaction
at threshold \cite{BFG97}. Radiative transitions seen in a typical X-ray
spectrum normally terminate at a circular atomic level $(n=l+1,l)$ having
a nonradiative, absorptive width of order 1-10 keV.
Lower atomic levels which are inaccessible  via the
atomic cascade process are expected
to be considerably broader, judging by the 2 - 3 orders of magnitude
increase in width going from  $(n+1,l+1)$ to  $(n,l)$. The interest
in observing such `deeply bound' atomic levels stems from anticipating
a larger overlap of the corresponding wave functions with the nuclear
density profile, and hence a greater sensitivity to the hadron-nucleus
strong interaction. However, once the width becomes of order 1 MeV, it
should be increasingly difficult to resolve these levels.

Todate, deeply bound hadronic atom levels have been directly observed
only for pions \cite{Yam96,Yam98},
using the ($d,^3$He) recoilless reaction \cite{HTY91}
on $^{208}$Pb, following earlier predictions that the $1s$ and $2p$
atomic levels in pionic Pb have widths of order 0.5 MeV, significantly
less than the approximately 1.5 MeV spacing \cite{FS85,Tok88}.
This striking narrowness is due to the well established repulsive $s$-wave
part of the pion-nucleus potential at threshold which pushes the
corresponding atomic wave functions out of the nucleus such that their
overlap with the nucleus, and hence with the imaginary part of the potential,
is substantially reduced. A similar, but not as favourable situation
might occur for $\Sigma^-$ atoms due to the inner repulsion of the
$\Sigma$ nucleus potential \cite{BFG94a,BFG94b,MFGJ95}. The other hadronic atom
species which have been studied experimentally, consisting of $K^-$ and
$\bar p$ atoms, do not appear at first sight likely candidates for
narrow deeply bound states, since the real part of the hadron-nucleus
potential at threshold is known for these hadronic species to be
strongly attractive and, furthermore, the imaginary (absorptive) part
is particularly strong, reaching (absolute) values of order 50 - 100
MeV inside nuclei \cite{BFG97}. These features of the $K^-$-nucleus
optical potential $V_{opt}$ follow qualitatively from various
microscopic studies of $V_{opt}$ utilizing the dominant effect of the
$\Lambda (1405)$ subthreshold unstable bound state
\cite{AHW76,Thi78,BWT78,MHT94,Koc94,WKW96,WRW97,Lut98}.

In this Letter we wish to point out, for strongly absorptive potentials,
a {\it new} mechanism for suppressing the widths of deeply bound atomic
states. It is based on the observation made by Krell long time
ago \cite{Kre71} 
that  absorptive potentials
generate  effective repulsion
which  pushes the wave function further outside the nucleus.
We demonstrate this effect for {\it realistic} $K^-$ nucleus optical
potentials obtained by fitting to a comprehensive set of $K^-$ atomic
levels reached in atomic cascade, as deduced from $K^-$ X-ray spectra.
Thus, we predict that $K^-$ deeply bound atomic states, generally, are
remarkably narrow and, therefore,
warrant experimental study.

The interaction of the $K^-$ meson at threshold with the nucleus
is described by the Klein-Gordon
 equation of the form:

\begin{equation}\label{equ:KG1}
\left[ \nabla^2  - 2{\mu}(B+V_{opt} + V_c) + (V_c+B)^2\right] \psi = 0~~ ~~
(\hbar = c = 1)
\end{equation}
where $\mu$ is the $K^-$ - nucleus reduced mass,
$B$ is the complex binding energy
 and $V_c$ is the
Coulomb interaction of the $K^-$ with the nucleus.
Only the leading term
in  $V_{opt}$ is retained in Eq. (\ref{equ:KG1}).
We have checked
that this neglect is well justified for atomic states.
The phenomenological density dependent (DD) potential of Friedman
{\it et al.} \cite{FGB93,FGB94} which generalizes and updates Batty's
earlier analysis \cite{Bat81} 
is given by:

\begin{equation}\label{equ:DD}
2\mu V_{opt}(r) =
 -{4\pi}(1+{\frac{\mu}{m}})b(\rho)
\rho(r) \;\;\; ,
\end{equation}

\begin{equation}\label{equ:b}
b(\rho)= b_0+B_0(\frac{\rho(r)}{\rho_0})^\alpha \;\;\; ,
\end{equation}
\noindent

\noindent
where $b_0$ and $B_0$ are complex parameters determined from fits to
the data, $m$ is the mass of the nucleon and $\rho(r)$
is the nuclear density distribution normalized to the number of nucleons
$A$, and $\rho_0=$0.16 fm$^{-3}$ is a typical central nuclear density.
`Macroscopic' nuclear densities \cite{BFG97} were used in most of the
present calculations.
For a detailed discussion of the DD potentials
as well as of the $\chi^2$ fits to the kaonic atom data see
Ref.~\cite{BFG97}. Here we note only that the full data base for kaonic
atoms, containing 65 data points over the whole of the periodic table,
was used in the fits, which yielded  $\chi^2$ per point
around 1.6, representing good fits to the data.
In this work, in order to test the dependence of our predictions on the type
of fitted potentials, we use two potentials of Table 6 in Ref. \cite{BFG97}.
The first potential is of a $t\rho$ form ($b_0=0.62+i0.92$~fm, $B_0=0$)
 with depths of about 70 MeV for the
(attractive) real part and 110 MeV for the (absorptive) imaginary
part. The second potential is of a DD form
satisfying the low density limit (with $b_0$ given by the $K^-N$ scattering
amplitude at threshold in free space). Its parameters
($b_0 = -0.15+i0.62$ fm
and $B_0 = 1.66 -i0.04$ fm, $\alpha$ = 0.24) produce a considerably stronger
attraction in the nuclear interior, but weaker absorptivity than for
the $t\rho$ potential.
Both give a good account of the $K^-$ atomic  data with
the DD potential  yielding a
significantly better fit. Parameters for the density distributions
$\rho (r)$ are given in Table 2 of Ref.\cite{FGB94}.

\begin{figure}
\epsfig{file=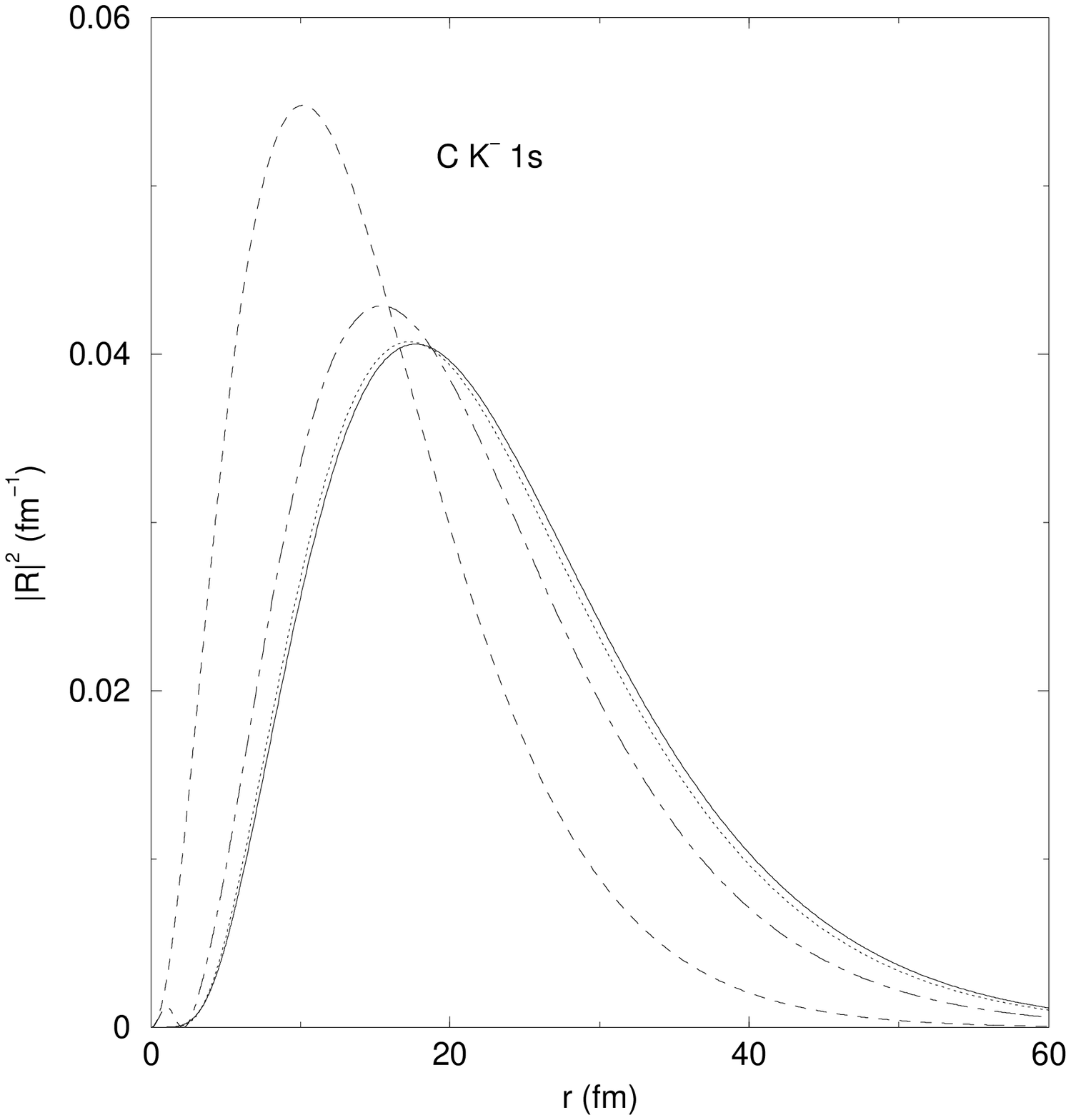,height=160mm,width=135mm,
bbllx=37,bblly=129,bburx=516,bbury=632}
\caption{Absolute value squared of the $1s$ radial wave function for
kaonic atoms of C. Dashed curve for electromagnetic interaction only,
solid curve with the full optical potential included, dotted curve
for only the imaginary potential included and dashed dot curve for only
the real optical potential included.}
\label{fig:C1s}
\end{figure}

We begin the discussion with the $1s$ atomic state in kaonic carbon
where a width of 62 keV is calculated together with a binding energy of
270 keV, which makes this level resolved from the observed 2 keV broad
$2p$ atomic level bound by 111 keV.
Figure \ref{fig:C1s} shows the absolute value squared of the radial
wave function of the  $1s$ state. The dashed  curve
is for the Coulomb interaction  due to the (finite size)
nuclear charge distribution and including also the usual vacuum
polarization terms. The other curves include additional components
of the $t\rho$ potential,
as obtained from fits to kaonic atom data.
The solid line is for the full complex
$t\rho$ potential added, the dotted line is
for adding only its imaginary part, and
the dashed-dot curve is for adding only its real part.
The results are almost indistinguishable if the DD
potential is used instead. A striking difference
compared to
pionic atoms \cite{FS85,Tok88}
is  immediately clear from this figure, namely, that the
strong interaction effects are completely dominated by the imaginary
potential. The strength of the imaginary potential is such that the
wave function is effectively excluded from the nuclear interior
($r {< \atop \sim} 3$ fm), and
the addition of the {\it attractive} real potential has a negligible
effect on the wave function. This repulsion reduces sufficiently the
overlap with the nucleus, and hence with Im $V_{opt}$,
such that the width of the $1s$ state becomes 50 times
smaller than would have been the case for the Coulomb wave
function (3.1 MeV).

A comment on the dash-dot curve in  Fig. \ref{fig:C1s} showing the
wave function for Im $V_{opt}$=0 is in order. This wave function
displays {\it repulsion} with respect to the Coulomb wave function
shown by the dashed curve, even though Re $V_{opt}$ is {\it attractive}.
This phenomenon occurs in hadronic atoms whenever Re $V_{opt}$ is deep
enough to bind {\it nuclear} states (\cite{Kre71}; see also \cite{GFB96}).
Since the real {\it atomic} wave function is orthogonal to the
wave functions of these deeper states, it develops nodes inside the
nucleus. Such a node is seen in Fig. \ref{fig:C1s}. These extra
nodes, and the actual $l$-value, determine the position of the main peak
of the atomic wave function, which in the present case occurs further
out than the peak of the Coulomb wave function, signifying repulsion.
When Im $V_{opt}$ is switched on to its full strength, it is clear
from Fig. \ref{fig:C1s} that the $K^-$ wave function gets thoroughly
suppressed in the nucleus. This suppression depends weakly  on
Re $V_{opt}$, which means that for strongly absorptive potentials
the nuclear states affect marginally the atomic states.

\begin{figure}
\epsfig{file=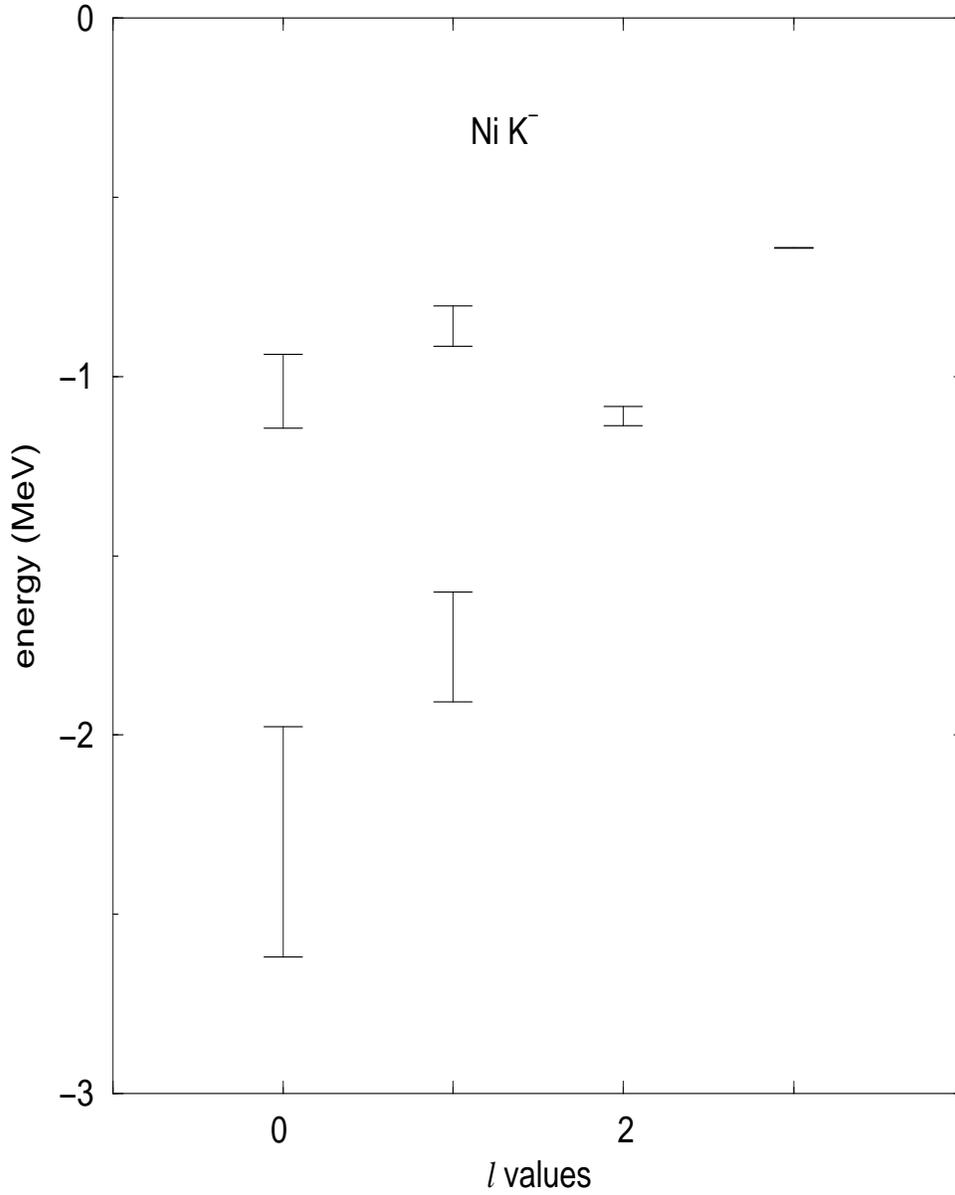,height=160mm,width=135mm,
bbllx=53,bblly=167,bburx=534,bbury=632}
\caption{Deeply bound levels in kaonic atoms of Ni
calculated using the $t \rho$ potential specified in the text. The  bars stand
for the full width  
$\Gamma$ (=2 Im $B$) of the levels and the centers
of the bars correspond to the energy ($-$ Re $B$).}
\label{fig:NiE}
\end{figure}

\begin{figure}
\epsfig{file=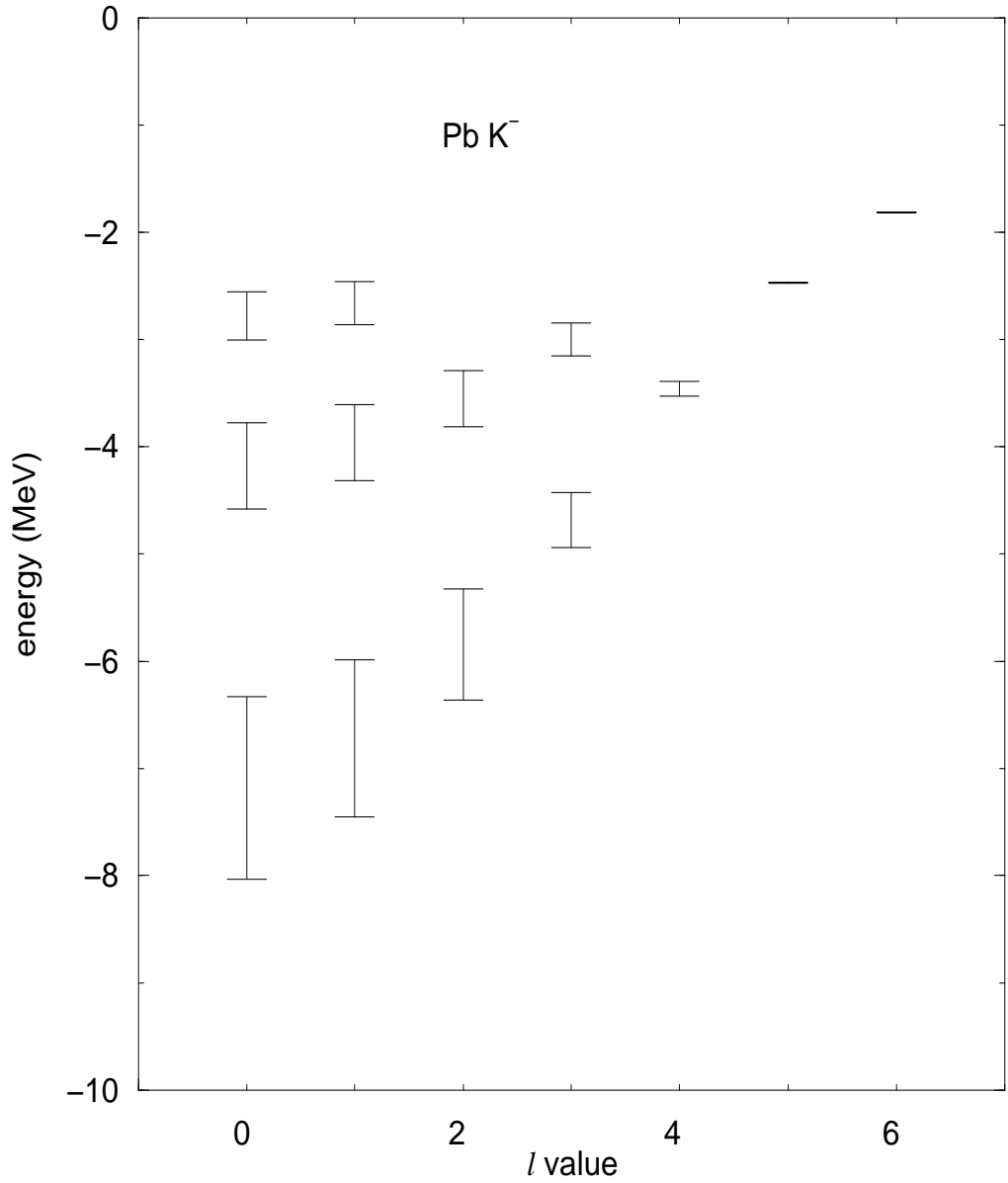,height=160mm,width=135mm,
bbllx=42,bblly=129,bburx=509,bbury=632}
\caption{Deeply bound levels in kaonic atoms of Pb (see 
caption of FIG. 2 for details).}
\label{fig:PbE}
\end{figure}

The phenomenon of well resolved deeply bound atomic states is not confined to
light kaonic atoms but is universal all over  the periodic table.
Figure \ref{fig:NiE} shows  the calculated `deeply bound' portion of
the $K^-$ atomic spectrum in Ni as a typical
medium weight nucleus and Fig. \ref{fig:PbE} shows similar results
for Pb. For these $N>Z$ nuclei we have used neutron densities
that slightly extend beyond those for protons, with $R_n-R_p$ = 0.20 fm
for Ni and 0.36 fm for Pb, where $R_n$ and $R_p$ are the half density
radius parameters. 
In both examples one sees many deep atomic levels whose widths
are less than 1 MeV. Although some overlap is observed in the
case of heavy nuclei,
we note that by using $l$-selective reactions to populate such states
certain levels could be expected to be reasonably well resolved.

\begin{figure}
\epsfig{file=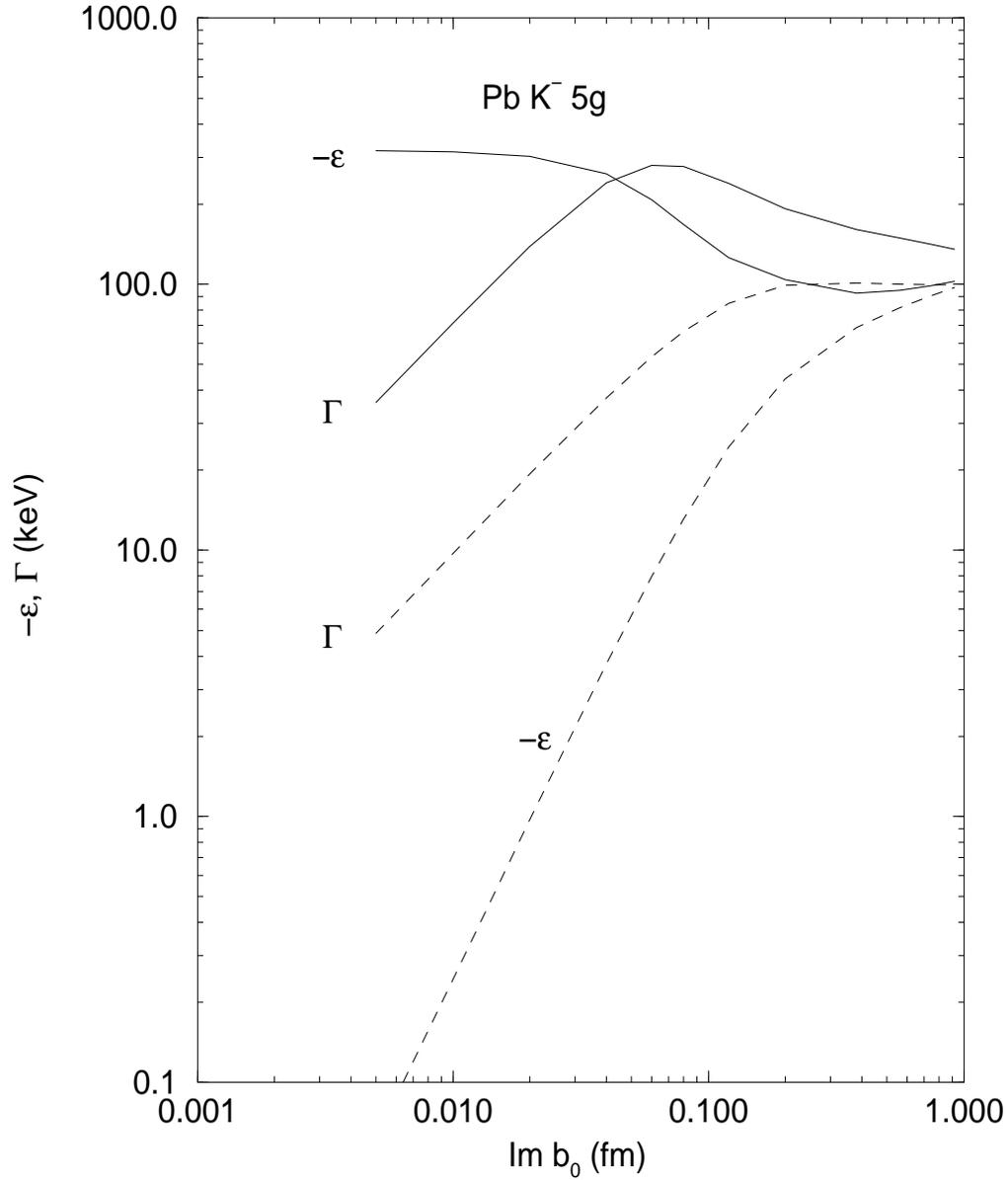,height=160mm,width=135mm,
bbllx=74,bblly=126,bburx=526,bbury=632}
\caption{Strong interaction shift and width for the $5g$ state in kaonic
atoms of Pb as function of the imaginary part of $b_0$. Solid curves for
the full optical potential, dashed curves for imaginary potential only.}
\label{fig:Pball}
\end{figure}

Further insight into the mechanism responsible for the width suppression
is provided by Fig. \ref{fig:Pball},
showing strong interaction widths and shifts for the
$5g$ level in kaonic Pb calculated as a function of Im $b_0$
for the $t\rho$
potential. The solid curves are for the full complex optical
potential $V_{opt}$,
whereas the dashed curves are for the imaginary part of the optical potential
only (Re $b_0$ = 0). The real $V_c$ potential is
included, of course, in all cases.
For small values of Im $b_0$, the effect of the attractive
Re $V_{opt}$ is to
increase the width by roughly a factor of 8 with respect to that calculated
when Re $V_{opt}$ is switched off. It is remarkable, however, that in each of
these cases the width saturates such that the difference between the two
calculated widths becomes less than 50\%.
 This saturation and near equality of the calculated widths for
the nominal value of Im $b_0$ is observed for other deeply bound states
with different values of $l$,
 regardless of
whether the solid  curve for $\Gamma$
lies above the dashed  curve for $\Gamma$ (as in Fig. \ref{fig:Pball})
or below it (as is the case, for example,
for the $4f$ state in Pb and for the $1s$ state in C discussed in
connection to Fig. \ref{fig:C1s}).
 For the dashed curve, it is
seen that the width saturates already at about 15\% of the nominal value of
Im~$b_0$, a feature which is due to the exclusion of the wave function from the
nuclear interior. The shift in this case (with Re $b_0$ = 0) is repulsive.
When the attractive Re $V_{opt}$ is included, the (solid curve) width reaches a
maximum at as little as 7\% of the nominal depth and then begins to decrease.
That is accompanied initially by a fairly constant repulsive shift which
eventually goes down, remaining nevertheless repulsive. 
We recall that a repulsive shift
due to an {\it attractive} real optical potential indicates
the existence of deeply bound  {\it nuclear}
states \cite{GFB96}. These states,
for the nominal value of Im $b_0$, are very broad
($\Gamma$ of order 50 MeV and more) and are not expected to be seen as well
defined states.

In order to check the sensitivity of the results to variations in the
model, we have
replaced the $t\rho$ optical
potential by the DD potential  and found that although the
spectrum of the deeply bound nuclear states varies considerably,  the
overall picture regarding the {\it atomic} states
remains essentially unchanged.
 For example, the widths of the circular $l$=1...5 levels
 in Pb change by less than 5\% when the DD potential is used.
Similar changes are observed in the calculated widths when the macroscopic
densities are replaced by single particle densities (see Ref. \cite{BFG97}).
The positions of the above levels change by less than 12 keV.
The largest sensitivity to the potential used and to the type of nuclear
density is observed for the binding energy of the  $1s$ level,
where the variation
 is up to 40 keV, being however only 2.5\% of the
width of the level.
We point out that the extrapolation from the atomic states observed in the
atomic cascade process, to the `deeply bound' atomic states, involves
6 MeV at most for Pb, and that no strong model dependence  is expected
over such an energy interval. In contrast, the location of the very deep and
broad {\it nuclear} states depends sensitively on the $V_{opt}$ used.

Before closing, we wish to mention a few candidate reactions
to form $K^-$ `deeply bound' atomic states. The lesson gained
from searching for such $\pi^-$ states \cite{Yam96,Yam98,HTY91}
is that a particularly
low momentum transfer in the forward direction,
say $q {< \atop \sim} 50$ MeV/c,
is necessary in order to achieve good angular momentum selectivity
(${\Delta}l \sim qR$) and to minimize the suppressive effects of the
nuclear distortion. This calls for using as low-energy $K^-$ beam
as possible, short of using stopped $K^-$. At the AGS, however, the
momentum transfer in the reaction ($K^-,p$) for example, with
a typical $K^-$ incident momentum $p_L$ = 600 MeV/c and for protons
emitted in the forward direction, is already too large to be useful
($q \sim 180$ MeV/c). Particularly low-energy $K^-$, of kinetic energy
$T_{K^-}$ = 16 MeV, can be produced in the decay at rest of the
$\phi$(1020) meson. The corresponding momentum transfer 
in the in-flight reaction
($K^-,p$), $q \sim 50$ MeV/c, is sufficiently small in this case,
by far smaller
for example than the value $q \sim 110$ MeV/c for the reaction ($K^-,\gamma$).
Such an  experiment could  be planned in the $e^+$-$e^-$ collider
$DA{\Phi}NE$, or by using quasi free $\bar pp$
 annihilation in a suitable
facility to produce slow $\phi$ mesons, as considered very recently by
members of the GSI collaboration
looking for meson-nuclear bound states \cite{GKY99}.
For a `beam' of $\phi$ mesons produced with momentum $p_L$ = 262 MeV/c,
 it is then possible
to inject strictly zero kinetic energy antikaons into a nuclear target
by  observing  $K^+$
in the forward direction with the same momentum in the recoilless
reaction  denoted by ($\phi,K^+$). 
 We defer a more detailed discussion
of this reaction to a separate publication.

In conclusion, we have shown that `deeply bound' kaonic atom states,
i.e. states which are inaccessible via the X-ray cascade process, are
sufficiently narrow to make them well resolved, particularly in $l$-selective
reactions of the type mentioned above.
The mechanism responsible for the suppression
of widths in the presence of very strong absorptive potentials is the
effective repulsion induced by such absorptivity. The net effect is repulsion
regardless of the presence of strongly attractive real potentials and,
in fact, the energy level spectra are dominated by the imaginary part of
the potential. The calculated `deeply bound' atomic spectra are remarkably
insensitive to the model used for the potential or to the model used for
the nuclear density, provided fits are made to the data on strong interaction
effects in `normal' kaonic atoms. Preliminary results suggest similar
effects in antiprotonic atoms.

\vspace{15mm}

We wish to thank A. Gillitzer, P. Kienle and T. Yamazaki for useful
communications.
This research was partially supported by the Israel Science Foundation.

\end{document}